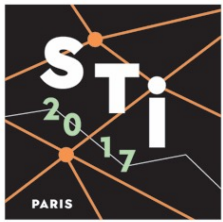

# Mapping knowledge with ontologies: the case of obesity

Montserrat Estañol[*], Francesco Masucci[**], Alessandro Mosca[***] and Ismael Ràfols[****]

[*] montserrat.estanyol@sirisacademic.com,
SIRIS Lab, Research Division of SIRIS Academic, Barcelona (Spain)
Dept d'Enginyeria de Serveis i Sistemes d'Informació, Universitat Politècnica de Catalunya, Barcelona (Spain)

[**] francesco.masucci@sirisacademic.com,
SIRIS Lab, Research Division of SIRIS Academic, Barcelona (Spain)

[***] a.mosca@sirisacademic.com,
SIRIS Lab, Research Division of SIRIS Academic, Barcelona (Spain)

[****] i.rafols@ingenio.upv.es
*Ingenio* (CSIC-UPV), Universitat Politècnica de València, València (Spain)

## ABSTRACT

Scientometric techniques have been remarkably successful at mapping science but they face important difficulties when mapping research for societal problems – possibly because they they are derived only from scientific documents and thus do not rely on non-academic expert knowledge. Here we aim to explore how ontologies can be used in science mapping, thus enriching current algorithmic techniques with systematic domain expert knowledge. This study introduces the methodology behind the construction of an ontology and tests potential uses in science mapping. We use *obesity* as a topic of case study.

## INTRODUCTION

Scientometric techniques have been remarkably successful at mapping science (Börner et al., 2003), with increasing capacity to produce fine-grained and apparently accurate descriptions of scientific topics (Klavans and Boyack, 2016). However, their use for science policy, in particular in relation to mapping research for societal problems, faces two important hurdles.

First, the validity of the maps has only been tested based in comparisons across algorithms using the same data (e.g. Velden et al., 2017), rather than in comparison with criteria derived from other sources (i.e. not based on bibliographic data). In particular, we are not aware of studies carrying out rigorous validations of scientometric maps on the basis of insights independently obtained from domain experts. Although experts may find specific maps reasonable and consistent with their intuitions (as often reported), no study has compared how experts responded to contrasting representations of science.

Second, since the data used in mapping is based on bibliographic data (titles, abstracts, keywords and references), the representations obtained reflect scientific discourse. This approach is appropriate for producing maps of the internal scientific dynamics, but it may be less useful to obtain maps that reflect problems addressed and solutions produced by research

− which is important when research aims to help address societal challenges, such as health conditions (obesity) or environmental problems (climate change). To reflect perspectives closer to those of practitioners, some exercises have produced maps based on publications but using health classifications (MeSH terms, Leydesdorff et al., 2012) or agricultural descriptors (based on CABI, Rafols et al., 2015) and semantic analyses based of other data sources, such as websites of NGOs (Klavans and Boyack, 2014), policy reports (Venturini et al., 2014) and parliamentary questions (Cassi et al., 2017).

Albeit useful, none of these methods uses explicitly domain knowledge to make or assess the maps or classifications. Also, it is poorly understood how the results of these science maps differ from expert-based perspectives on the research topics examined. One must keep in mind as well, that when addressing societal challenges, expert perspectives may differ significantly depending on the framings adopted particularly in multi- and transdisciplinary contexts (Lawrence, 2004). As a result, it is crucial for informing policy to provide representations that are comprehensive and encompass plural perspectives. One way forward is to contrast or complement scientometrics (or algorithmic) maps with 'mappings' obtained from other methods based on expert insights, such as interviews, focus groups or surveys.

Another alternative that we explore here is the use of domain ontologies. An *ontology is a formal model representing expert domain knowledge through a set of concepts and relationships between them*. The domain knowledge is gathered by interviews and document, and it is specified in a given formal language. Being a mathematical logic artefact, algorithms (theorem provers) can then be used to infer 'new knowledge', that is implicit in the initial model.

In this study, we aim to explore how ontologies can be used in science mapping, thus enriching current algorithmic techniques with systematic domain expert knowledge. We envisage that ontologies can be used for the following purposes:
1. create a representation of the topic, given that an ontology is formally a mathematical graph (a network), and thus be visualised.
2. tag or label current science mapping approaches, thus characterising clusters or network zones.
3. impose constraints in current science mapping methods, i.e. in conventional methods for classifying entities into topics, such as co-citation clustering or topic modelling.

This study introduces the methodology behind the construction of an OWL[1] (Ontology Web Language) ontology and tests potential uses. We use *obesity* as a topic of case study. Obesity is a multifactorial disease and it is considered to be one of the greatest epidemics of the current century.[2] As any disease (or problem) obesity can be studied from a variety of perspectives, including the diverse causes, mechanisms, consequences, prevention, diagnostic, therapeutics and public health interventions.

The construction of the obesity ontology is framed within a larger goal that aims at going beyond the state of the art in statistical modelling for data classification by imposing logical constraints to topic modelling (following attempts by Andrzejewski et al., 2009). This would allow to systematically compare science maps resulting from unsupervised (purely algorithmic) methods and with those obtained with knowledge-enhanced methods (enriched by the domain knowledge embedded in the ontology).

---

[1] https://www.w3.org/OWL/
[2] "WHO | Obesity - World Health Organization." http://www.who.int/topics/obesity/en/. Accessed 21 Mar. 2017.

## PREVIOUS ONTOLOGIES FOR OBESITY

To our knowledge, the present ontology is the first attempt in the literature to cover the topic of obesity from a wide array of perspectives. As detailed below, previous works on obesity were either *ad-ho*c to specific services or their purpose was more specific, and thus they were not appropriate to meet the goals of our exercise which are more comprehensive and plural.

Sojic, Terkaj & Contini (2016) present a modular ontology whose purpose is to capture individual profiles of people and infer information about their general health state (including eating habits, their sleep quality, and their physical activity) - in terms of obesity-related aspects. For example, given certain information of a person - such as their birthdate, their body-mass index (BMI), their eating habits, etc. it is possible to exploit their ontology to determine if the person suffers from a disease (such as hypertension, obesity, etc.), by means of reasoner.

Scala et al. (2012) focused on diagnosing obesity and its comorbidities given certain information about a person. Like Sojic, Terkaj & Contini (2016), they created an OWL ontology and a set of rules, formally specified in SWRL[3] (Semantic Web Rule Language) standard, with the aim of automatize the diagnostic process. The work of Sojic, Terkaj & Contini (2016) covers a wider variety of areas - not only diseases.

Kim et al. (2013) developed an ontology for a mobile application with the goal of helping users make the necessary lifestyle changes to lose weight. The application is able to infer information about the health state of an individual given certain parameters, and offer recommendations. Unfortunately, it is not possible to compare it with the other approaches as the ontology is not publicly accessible.

Shaban-Nejad, Buckeridge & Dubé (2011) created an ontology (COPE) to promote healthy habits and disseminate knowledge about obesity, focusing specifically on childhood obesity. This ontology integrates various perspectives such as food and nutrition, diseases, social and environmental factors, behavioural parameters and media. Unfortunately, we could not evaluate and reuse the ontology itself as we could not get access to it.

Finally, Elhefny, Elmogy & Elfetouh (2014) define an ontology to represent the different types of cancer related to obesity and its diagnosis. Their main purpose is to share knowledge, reuse it and reason on it. The ontology, which contains 60 classes (i.e. concepts related with the issue, such as 'illness') and 62 instances (i.e. practical cases of those concepts, such as 'hypertension'), does not seem to be fully developed, and the classification of some concepts is questionable.

Note that, except for the work of Shaban-Nejad, Buckeridge & Dubé (2011), the ontologies reviewed here had a completely different purpose from ours - evaluating the health state of a person (Sojic, Terkaj & Contini, 2016; Scala et al., 2012; Kim et al., 2013) or describing the different cancers linked to obesity (Elhefny, Elmogy & Elfetouh, 2014). Because of this, they are too much focused on specific goals to be considered as suitable representations of the multifactorial nature of obesity. In the case of the COPE ontology (Shaban-Nejad, Buckeridge & Dubé, 2011), lack of access to it made it impossible to determine if it was reusable to some extent.

---

[3] https://www.w3.org/Submission/SWRL

## AN ONTOLOGY FOR OBESITY

For the creation of the obesity ontology, we have studied a corpus of literature from multiple sources, ranging from introductory Wikipedia pages and medical websites, to scientific articles from various disciplines, such as medicines or social science. In addition, four obesity experts in science public health were interview for specific knowledge acquisition sessions, with the aim of capturing diverse current discourses on obesity. We also checked online terminologies, such as AGROVOC, medical websites and World Health Organization (WHO) documents. Note that the searches have not been limited to medical perspectives. Since all experts agree on the fact that obesity is a *multifactorial disease*, we have also considered documents taking a variety of perspectives, including marketing and its impact on food consumption, lifestyles, psychology, and nutrition, among others.

On the basis of the literature analysis, the current version of the ontology contains around 150 concepts (i.e. classes), 40 relationships between concepts (i.e. object and data properties), and more than 500 instances (i.e. individuals). Figure 1 shows a screenshot of the current statistics of the ontology, viewed using Protégé.[4]

**Figure 1: Screenshot showing the Obesity ontology statistics in Protégé, highlighting the number of classes, number of object and data properties and the quantity of individuals.**

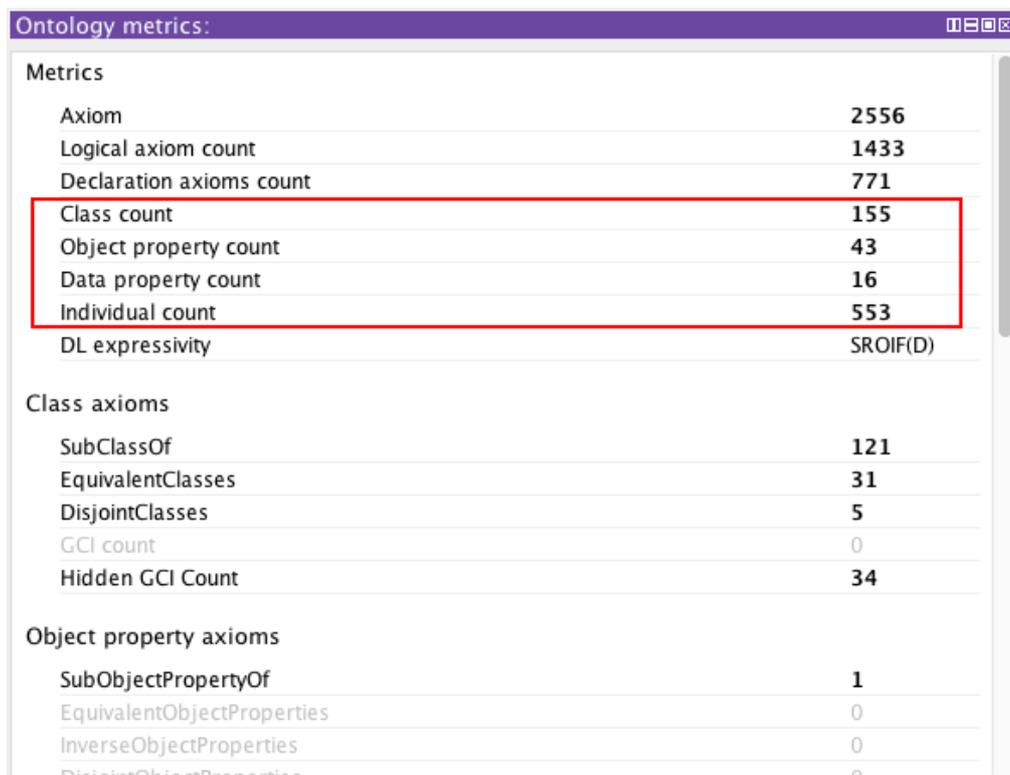

The key concept in the obesity ontology is that of *Disease*. *Obesity* is an individual of (or, an 'instance of') *Disease*. As shown in Figure 2, the concept of *Disease* is part of a hierarchy (taxonomy) of concepts. More specifically, a *Disease* is a *PathologicalCondition*, and a *PathologicalCondition* is, in turn, a *MedicalCondition*. A *MedicalCondition*, which represents any condition or state that may require medical treatment, may have a set of *Manifestations*, represented through relationship *medConHasManifestation*. In turn, a *Manifestation* can either be a *MedicalSign* or a *Symptom*. For example, a *Hernia* has *AbdominalPain* (a *Symptom*) as a *Manifestation*.

---

[4] http://protege.stanford.edu

A *MedicalCondition* may also lead to other *MedicalConditions* (e.g., *Obesity* can lead to *Type2Diabetes*, or *NightEatingSyndrome*, a psychological disorder, may lead to *Obesity*). This is represented using relationship *medCondMayLeadToMedCond*.

**Figure 2. Screenshot of the UML class diagram representing the taxonomy of concepts related to the *Disease* concept.**

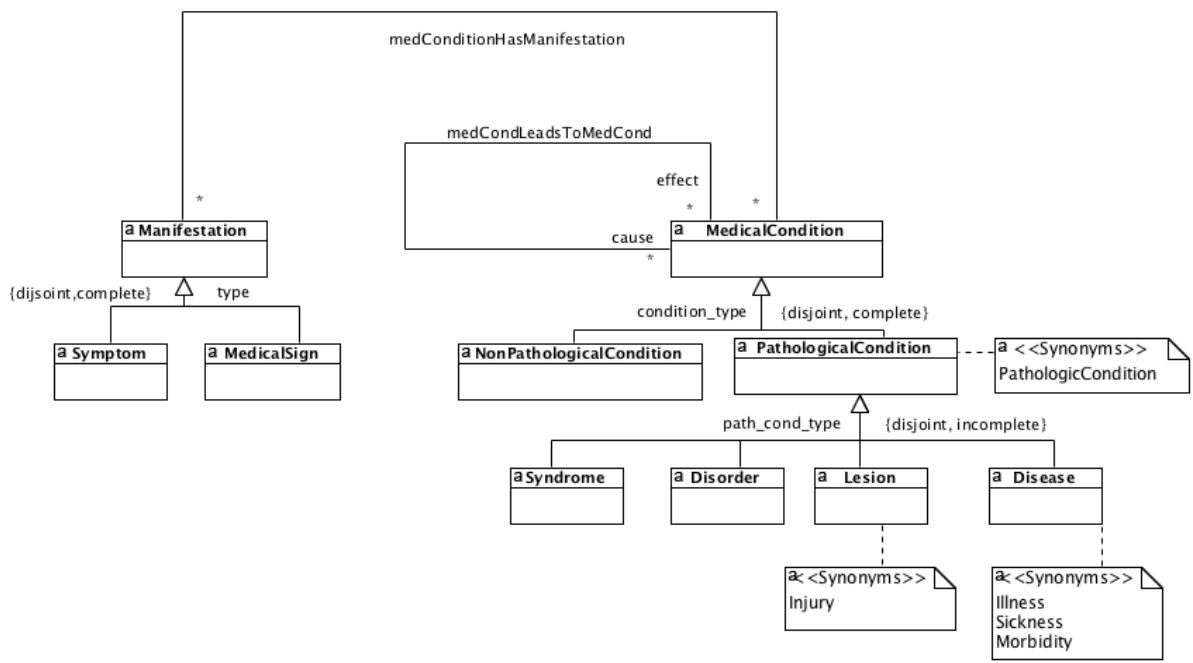

As it can be seen in Figure 3, *MedicalConditions* can be dealt with by following a *Treatment*, represented through the relationship *canBeTreated*. Treatment is a concept that groups the following concepts: *Medication* (e.g., *Liraglutide*), *Behaviour* (e.g., *PhysicalActivity*), *Diet* (e.g., *Low-CalorieDiet*), *MedicalProcedure* (e.g. *Roux-en-YGastricBypass*) and *Therapy* (e.g., *BehaviouralTherapy*). For example, *Obesity* can be treated by administering *Liraglutide*, by carrying out *PhysicalActivity*, by eating a *Low-CalorieDiet*, by undergoing *Roux-en-YGastricBypass* surgery and/or by attending *BehaviouralTherapy* sessions.

**Figure 3. Screenshot of the UML class diagram representing *Treatment* and its relationship to *MedicalCondition*.**

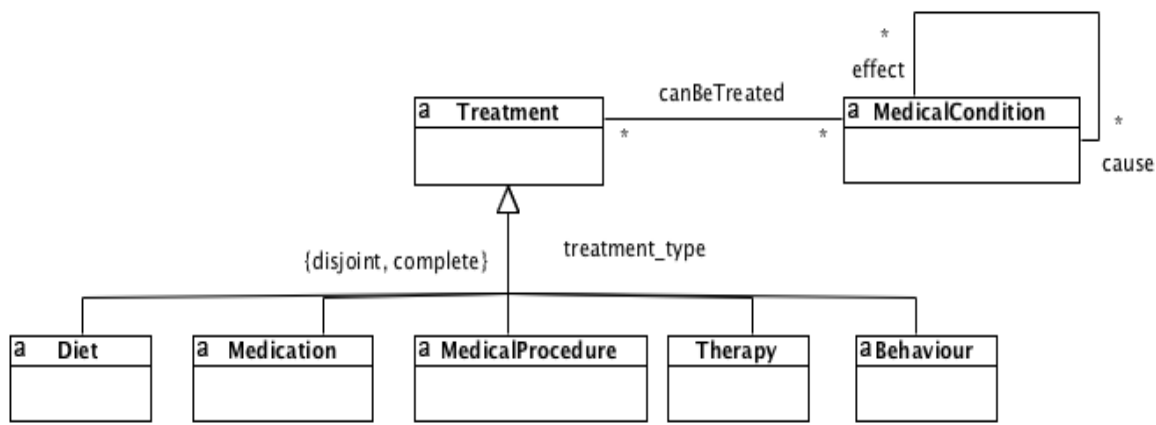

However, not only *MedicalConditions* can lead to other *MedicalConditions*. Any kind of *Treatment* (which may include taking a *Medication*, following a certain *Behaviour* or *Diet*, or undergoing a *MedicalProcedure* or *Therapy*) may have side effects, resulting in a certain *MedicalCondition*. For example, *Liraglutide*, a *Medication* used for the treatment of obesity, can cause *Hypoglycemia* (low blood sugar). Similarly, *Environmental* and/or *Personal Circumstances*, such as *Overeating* or a *SedentaryLifestyle* may also contribute to a *Disease*.

While these examples come from a medical perspective, the ontology covers other perspectives, such as the nutritional one. For instance, it includes a taxonomy of *Nutrients* (see Figure 4) which represents concepts such as *Carbohydrate*, *Fat* or *Protein*, to mention a few examples. A *Diet* restricts certain types of *Nutrient*, e.g., a *Low-CarbohydrateDiet* will restrict the amount of *AvailableCarbohydrates* that a person can eat; similarly, a *Low-FatDiet* will limit the daily *Fat* intake. Both diets can be used as a treatment for *Obesity*.

**Figure 4. UML class diagram representing the taxonomy of Nutrients.**

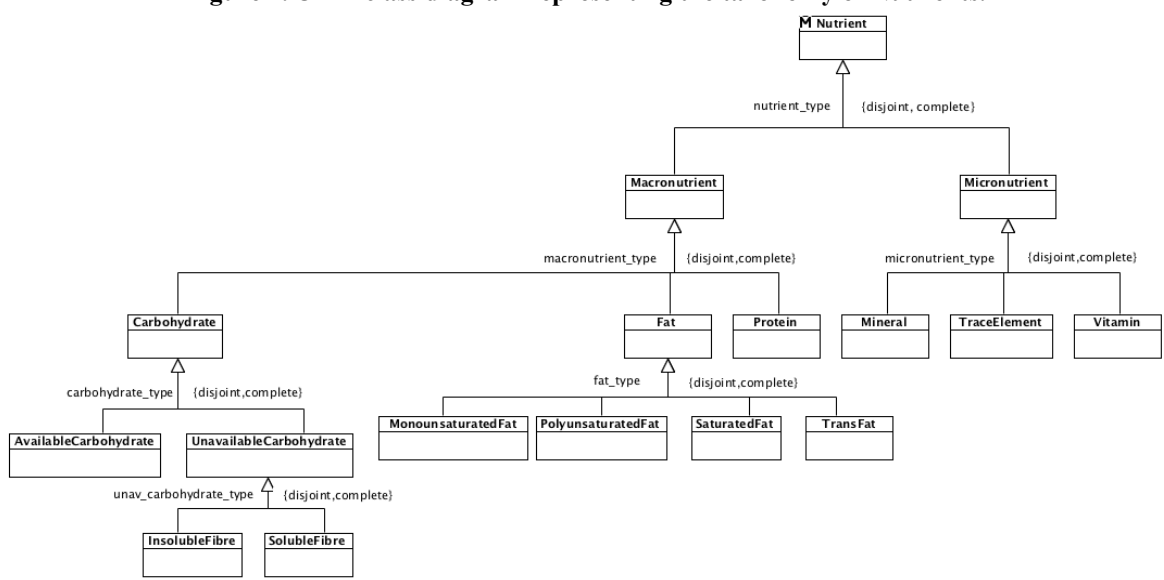

Apart from representing the concepts and their relationships, the expressivity of the formal language allows us to establish constraints (by means of logical axioms) over them. For example, each object property in the ontology may have a *domain* and a *range* specified. The domain and range constrain the type of the 'source' and the 'target' of a given binary relation in the ontology. The domain represents the concepts over which the relationship is defined, the range represents the target concept(s) of the relationship. In the case of relationship *canBeTreated*, mentioned earlier, the domain is *MedicalCondition* and the range *Treatment*. We define domain and range for every relationship in our ontology. Notice that the presence of domain-range constraints, beside the documentation-oriented goal, is relevant at the moment of using the ontology in combination with a theorem prover (or reasoner, i.e., an algorithm which is able to automatically infer implicit knowledge from the ontology). The reasoner is then able to highlight, for instance, that a certain individual has been wrongly added into the ontology, if it is the first argument of a relation and it does not comply with the domain constraint associated to that relation.

There are other axioms that can be defined over the relationships, such as symmetry, asymmetry, ir/reflexivity, in/transitivity, etc. Where appropriate, we use them to characterise the properties of the relationships. As we have seen, the relationship *medCondMayLeadToMedCond* represents that a *MedicalCondition* can lead to another *MedicalCondition*. Notice that in this case, both domain and range correspond to the same concept, *MedicalCondition*. As it does not make sense for a certain *MedicalCondition* to be caused by itself, the relationship requires the irreflexive axiom.

The last type of axiom we would like to point out here, is represented by the disjoint union of concepts. This kind of axiom is used to define a concept as the result of the (set-theoretical) union of various concepts which are pairwise disjoint. In our example, *Treatment* is actually the disjoint union of *Diet*, *Medication*, *Behaviour*, *Therapy* and *MedicalProcedure*. An instance cannot be both a *Diet* and *Medication*, or a *Diet* and a *Behaviour*, or a *Medication* and a *Therapy*, and so on.

Figure 5 shows the current list of classes or concepts in the ontology and their object properties (relationships).

**Figure 5. List of class and object properties (relationships) in the ontology.**

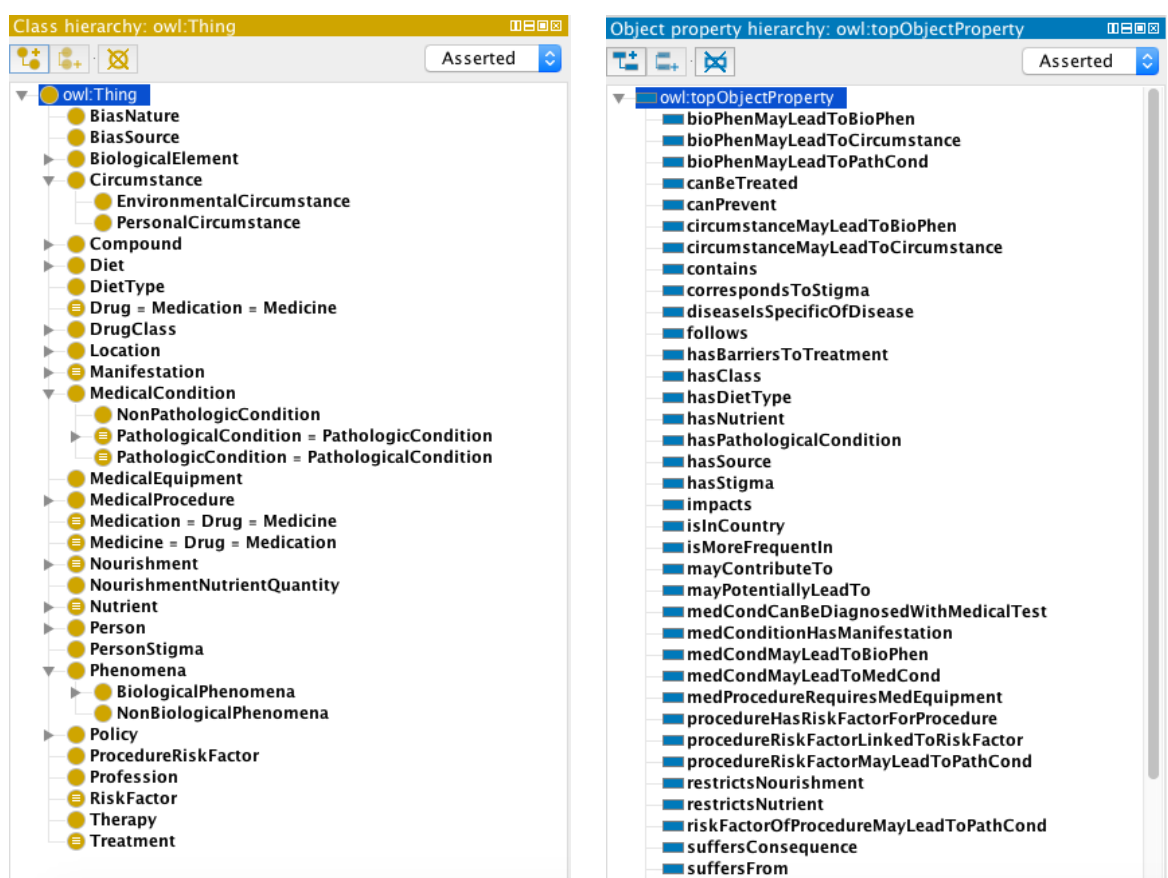

**PRELIMINARY RESULTS**

Figure 6 presents the network visualisation of the obesity ontology. While still very preliminary, it provides some insights: there is an area related to *diseases* partly caused by obesity (left side, e.g. diabetes, hypertension), an area related to *metabolism* (top side), and one area related to *environmental and behavioural causes* (including individual and social). In the presentation, we plan to elaborate this analysis in comparison to other science maps (e.g. Cassi et al., 2017).

**CONCLUSIONS**

To sum up, in this paper we have presented the creation of an ontology on the topic of obesity. with the final goal of using it as a structured representation of knowledge to aid in semantically-enhanced topic modelling. Despite the existence of other ontologies on the topic, lack of access or of appropriateness meant that they could not be reused. Moreover, as obesity is a multifactorial disease, the ontology covers the wide variety of perspectives involved in it, including the medical and nutritional perspectives, environmental factors, or physical activity, to mention a few examples.

As mentioned in the introduction, the creation of the obesity ontology is framed within the goal to compare the results obtained by unsupervised topic modelling and knowledge-enhanced topic modelling, the latter performed via an effective coupling of the ontology and LDA algorithm. Further work will focus on carrying out the tasks required by these two goals.

**Figure 6: The Obesity ontology, visualised as a graph. Each node is a logical concept related with obesity. Links are logical relations among concepts. Different colors denote different clusters of tightly related concepts.**

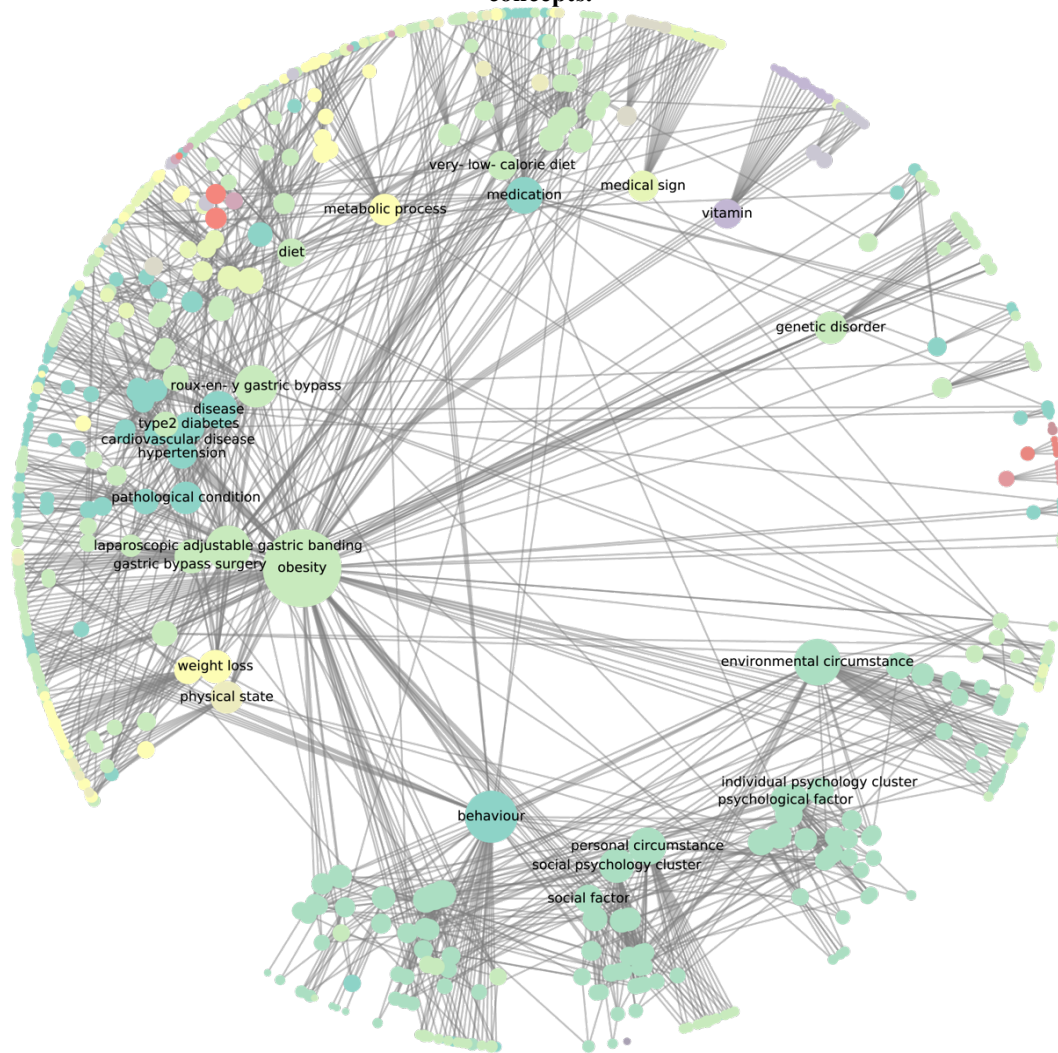